\documentstyle[12pt]{article}
\def\beq{\begin{equation}}
\def\eeq{\end{equation}}
\def\bea{\begin{eqnarray}}
\def\eea{\end{eqnarray}}

\def\ba{\begin{array}}
\def\ea{\end{array}}
\def\bce{\begin{center}}
\def\ece{\end{center}}

\begin{document}
\begin{titlepage}
\rightline{SNUTP-98-031}
\rightline{revised, Aug., 1998, hep-th/9804093}
\def\today{\ifcase\month\or
January\or February\or March\or April\or May\or June\or
July\or August\or September\or October\or November\or December\fi,
\number\year}
\vskip 1cm
\centerline{\Large \bf  Orbifolds of $AdS_7 \times S^4$ and 
Six Dimensional  $(0, 1)$ SCFT}
\vskip 1cm
\centerline{\sc Changhyun Ahn$^{a}$, 
Kyungho Oh$^{b}$ and Radu Tatar$^{c}$}
\vskip 1cm
\centerline{{\it $^a$ Dept. of Physics, Seoul National University,
Seoul 151-742, Korea}}
\centerline{{\tt chahn@spin.snu.ac.kr}}
\centerline{{ \it $^b$ Dept. of Mathematics, University of Missouri-St. Louis,
St. Louis, MO 63121, USA}}
\centerline{{\tt oh@arch.umsl.edu}}
\centerline{{\it $^c$ Dept. of Physics, University of Miami,
Coral Gables, FL 33146, USA}}
\centerline{{\tt tatar@phyvax.ir.miami.edu}}
\vskip 2cm
\centerline{\sc Abstract}
\vskip 0.2in
We analyze the relation between the large $N$ limit of 6 dimensional 
superconformal 
field theories
with eight supercharges and M theory on orbifolds of $AdS_7 \times {\bf S^4}$. 
We use the known spectrum of Kaluza-Klein harmonics of supergravity on $AdS_7 
\times {\bf S^4}$ and we take their orbifold projection 
to get information about 
the chiral primary operators of 6 
dimensional SCFT which is realized on the worldvolume of M5 brane sitting 
at the orbifold 
singularities.
\vskip 2in
\end{titlepage}
\newpage
\setcounter{equation}{0}


\section{Introduction}
\setcounter{equation}{0}

Recently, the connection between anti-de Sitter (AdS) space and the
dynamics of worldvolume of D brane has been actively studied. 
Maldacena~ \cite{mal} proposed  that
the large $N$ limit of super conformal field theories (SCFT) 
can be described by taking the
supergravity limit of the superstring compactified on AdS space.
Furthermore, the correlation functions in the SCFT that has 
the ADS boundary at infinity boundary as its spacetime can be 
obtained by taking the dependence
of the supergravity action on the asymptotic behavior of its fields at the
boundary \cite{polyakov,witten}.
This way one can get the scaling dimensions of operators in the SCFT from the 
masses of particles in string/M theory.

Moreover, for any given ${\cal N} =2, 1, 0$ superconformal model in 
4 dimensions
there exists a formulation of type IIB theory 
compactified on orbifolds of $AdS_{5} \times {\bf S^5}$ 
\cite{kachru,vafa}. This proposed duality has been tested by studying
the Kaluza-Klein (KK) states of theory on the orbifolds of
$AdS_5 \times {\bf S^5}$ and by comparing them with 
the chiral primary operators
of the SCFT on the boundary \cite{ot}.

As we go one step further, the field theory/ M theory duality gives
M theory on the $AdS_4$ or $AdS_7$ for some superconformal theory
in 3 and 6 dimensions, respectively. The maximally supersymmetric theories
in 3 and 6 dimensions have been studied recently 
\cite{aoy,lr,minwalla2,halyo1,gomis}.
The lower supersymmetric case was realized on the worldvolume of M theory
branes sitting on the orbifold singularities \cite{fkpz} (See also 
\cite{berkooz}).
Very recently, along the line of \cite{ot}, 
the Kaluza-Klein states of M theory on the orbifolds of
$AdS_4 \times {\bf S^7}$ were studied and compared 
with the chiral primary operators
of the SCFT on the boundary \cite{eg}.

In this paper, we investigate the Kaluza-Klein states of M theory
for an orbifolds of 
$AdS_7 \times {\bf S^4}$ and
we obtain the dimensions of operators in the superconformal
multiplets by using the correspondence between AdS compactifications and SCFT.
We obtain results refering only to the untwisted sectors of M theory
compactified on AdS space which is the one that can be derived from 
supergravity.  By taking all the degrees of freedom of M theory 
the singular geometry is regulated.


\section{ Review on $AdS_7$ and $(0, 2) $ SCFT }
\setcounter{equation}{0}

We briefly review the SCFT/$AdS_7$ relation
proposed in \cite{polyakov,witten}.
The boundary of $AdS_7$ is a
6 dimensional Minkowski space with points at infinity added.  The
isometry group of $AdS_7$ is $SO(6,2)$ which  is also the conformal
group on the boundary.  The proposed duality relates
M theory on $AdS_7$ to the large $N$ limit of SCFT on its boundary.  
In the Euclidean version the boundary is 6 sphere ${ \bf S}^6$.
Consider the maximally supersymmetric case, so that the
internal space is also a sphere.  Let $\phi$
be a scalar field on $AdS_7$ and $\phi_0$ its restriction to the
boundary ${ \bf S}^6$.  According to the SCFT/$AdS_7$ correspondence, $\phi_0$
couples to a conformal operator ${\cal O}$ on the
boundary through a coupling $\int_{{\bf S^6}} \phi_0 {\cal O}$.

When $\phi$ has mass $m$, the corresponding operator
${\cal O}$ has the conformal dimension $\Delta$ given by the following
relation
\bea
m^2 = \Delta(\Delta-6).
\label{dim}
\eea
The irrelevant, marginal and relevant operators of the boundary theory
correspond to massive, massless and tachyonic modes in the supergravity
theory.  If a $p$-form $C$ on $AdS_7$ couples to a $6-p$ form
operator $\cal O$ on the boundary through a coupling $ 
\int_{{\bf S}^6} C \wedge {\cal O}$, then the relation between the mass of
$C$ and the conformal dimension of $\cal O$ is given by
\bea
m^2 = (\Delta+p)(\Delta+p-6). 
\label{dimp}
\eea
The value of $m^2$ in this formula is the eigenvalue of the Laplace
operator on the $AdS_7$ space in the supergravity side. 
The values 
usually quoted for $p$-forms are the eigenvalues ${\widetilde m}^2$ of the
appropriate Maxwell-like operators. Their relation with the dimensions
is given by \cite{ffz,aoy}
\beq
{\widetilde m}^2 = (\Delta-p)(\Delta+p-6).
\label{dimpm}
\eeq
The massless graviton in the $AdS_7$ supergravity  couples
to the dimension $6$ stress-energy tensor of the SCFT.
When the internal space has continuous
rotational symmetry, there exist also $AdS_7$ massless vector
fields in its adjoint representation which  can couple to the five dimensional 
R symmetry currents of the SCFT.

Let us consider M theory on $AdS_7 \times {\bf S^4}$ with a 4-form flux of $N$
quanta on ${ \bf S}^4$, and with the radii of the 
$AdS_7$ and ${ \bf S}^4 $ being
$R_{AdS_7} = 2R_{S^4} = 2 l_p(\pi N)^{1/3}$. The eleven dimensional
supergravity is appropriate at energies much smaller than the Planck
scale $1/l_p$. For large $N$ this includes the energy range of the 
Kaluza-Klein
modes, whose mass is of the order of $1/R_{AdS_7}$. 
The bosonic symmetry of this
compactification of eleven dimensional supergravity is $SO(6,2) \times
SO(5)$.
In \cite{mal} it was proposed that the $(0, 2)$ conformal theory, which
is the decoupled theory on $N$ parallel M5
branes, is dual to M theory on the above background.  
The $SO(6,2)$ part of the symmetry of the
supergravity theory is the conformal group of the SCFT, which can be
thought of as living on the boundary of the $AdS_7$ space.  The $SO(5)$
part of the symmetry corresponds to the R symmetry of the
superconformal theory.

The Kaluza-Klein excitations of supergravity, in the maximally
supersymmetric cases fall
into small representations of supersymmetry. 
Thus, their mass formula is protected
from quantum and string/M theory corrections.  According to the proposal
in \cite{witten}, they can couple to
chiral fields of the SCFT on the boundary, for which
the scaling dimensions are also protected from
quantum corrections. 

The spectrum of the Kaluza-Klein harmonics of
M theory on $AdS_7 \times {\bf S^4}$ was
analyzed in \cite{van}. 
There are three families of scalar
excitations. Two families contain states with only positive $m^2$ and
which only correspond to irrelevant operators.  One family has also states
with negative and zero $m^2$.  They fall into
the $k$-th order symmetric traceless
representation of $SO(5)$ with unit multiplicity.
Their masses are \footnote{The
masses should be scaled by a factor of $e=2$.}
\bea
m^2 = 4k(k-3), \;\;\;\; k=2,3, \cdots
\label{scalarmass}
\eea
The field corresponding to $k=1$ also appears in the supergravity, and
this is the singleton which may be gauged away everywhere 
except at the boundary
of the $AdS$ space and decouples from all other operators. 
By using (\ref{dim}), the dimensions of the corresponding
operators in the SCFT are
\beq
\Delta = 2k, \;\;\;\; k=2,3, \cdots 
\label{dims}
\eeq
These are precisely the dimensions of the chiral primary operators
found in \cite{abs}, which parameterize the space of flat directions
$({ \bf R^5})^N/S_N$ in a gauge invariant way and 
this can be viewed as a test of the conjecture in \cite{mal}.

There is one family of vector
bosons that also contains massless states
\bea
{\widetilde m}^2 = 4(k^2-1), \;\;\;\; k=1,2, \cdots
\label{vectormass}
\eea
By using (\ref{dimpm}), the dimensions of the corresponding
1-form operators in the SCFT are given by
\bea
\Delta = 2k+3, \;\;\;\; k=1,2, \cdots 
\label{dimsp}
\eea
The massless vector obtained for 
$k=1$ in (\ref{vectormass}) corresponds to the 
dimension 5 R symmetry current in SCFT.
In general, chiral fields corresponding to all the towers of
Kaluza-Klein harmonics are related to the scalar operators of
(\ref{dims}) by the superconformal algebra, as discussed in the four
dimensional case in \cite{ffz}. Each value of $k$ gives rise (at least
for large enough $k$) to one field in each tower of KK states, with an
$SO(5)$ representation that is determined by the representation of the
scalar field. In particular, the R symmetry currents and the stress
energy tensor sit in the same superconformal representation as the
scalar field with $k=2$ mentioned above. 


\section{ An Orbifold of $AdS_7 \times S^4$ and $(0, 1)$ SCFT  }
\setcounter{equation}{0}

It is proposed that the $(0, 1)$ SCFT on $N$ M5 branes can be constructed
by investigating M theory on the orbifold of $AdS_7 \times {\bf S^4}$.
The result of \cite{fkpz} states that 
M theory on $AdS_7 \times {\bf S^4}/{\bf Z}_k$
is dual to $(0, 1)$ six dimensional theory with the gauge group
$SU(k)^{N-1}$ and with one tensor multiplet and k hypermultiplets in the
${\bf k} \oplus \bar{\bf k}$ representation of the gauge group factors.
The theory has a $SU(2)_R$ symmetry, the only bosons which are transforming
non-trivially under this symmetry being the ones in the hypermultiplets
and these transform in the fundamental representation.
For the M5-brane, the transverse space is ${\bf R}^5$ and an appropriate
orbifolding leads to $(0, 1)$ theories on the world-volume of the
M5-brane. In order to keep the $AdS_7$ structure in the near-horizon
geometry, the orbifolding should act on ${\bf S}^4$ only \cite{fkpz}.

One approach to obtain such orbifolds would be
to break ${\bf S}^4$ into fiberations of smaller dimensional spheres over
a base manifold.  Then one can obtain an orbifold by  taking actions on
the base. The $SO(5)$ isometry of ${\bf S^4}$ will be broken into
the isometry groups of both the fiber and the base. By embedding a discrete
group into the isometry of the base corresponding to the orbifolding process,
the isometry group of the base will be further broken.
This method has been inspired by  the work of
Entin and Gomis \cite{eg} in the context of 
$AdS_4 \times {\bf S^7}$.

To explain this explicitly, let
\bea
{\bf S}^4 &=& \{(x_1, x_2, x_3, x_4, x_5) \in {\bf R}^5 \,|\,
 x_1^2 + x_2^2 + x_3^2 + x_4^2 + x_5^2 = \rho\}\\ 
{\bf D}^i &=& \{(x_1, \cdots , x_{i}) \in {\bf R}^i \,|\,
x_1^2 + \cdots + x_i^2 \leq \rho \}
\eea
where $\rho$ is the radius of the 4 sphere ${\bf S^4}$ given by
$\rho = R_{S^4} = l_p (\pi N)^{1/3}$ as we mentioned previous section.
We list the possible fiberations of 
 ${\bf S}^4$ induced by the coordinate projections.

$\bullet$  Case 1
\bea
\pi_1: {\bf S}^4 \rightarrow {\bf D}^1,\quad 
(x_1, x_2, x_3, x_4, x_5) \rightarrow (x_1)
\eea 
Then the generic fiber will be a 
3 sphere ${\bf S}^3$ and the only isometry of the base 
leaving the boundary of the base invariant is ${\bf Z}_2$ whose action
is given by $x_1 \leftrightarrow - x_1$. 
Now we take the orbifold projection by ${\bf Z}_2$. Then we obtain 
the hemisphere of Berkooz~\cite{berkooz} which is made of a bundle
of 3-sphere over a 1 dimensional manifold ${\bf D}^1/{\bf Z}_2$.
Thus this theory has an $SO(4)$ symmetry. Furthermore, 
it is states in \cite{berkooz} that the field theory
can acquire a global
$E_8$ symmetry by placing $E_8$ vector gauge field on 
the fiber over $x_1 = 0$. 

$\bullet$ Case 2
\bea
\label{2}
\pi_2: {\bf S}^4 \rightarrow {\bf D}^2,\quad 
(x_1, x_2, x_3, x_4, x_5) \rightarrow (x_1, x_2)
\eea 
The generic fiber is 2 sphere ${\bf S}^2$ and the the base ${\bf D}^2$ 
can be embedded
 into
a complex $z$-plane ${\bf C}$ by letting $z = x_1 + ix_2$. Now the 
discrete subgroup ${\bf Z}_k$ of $U(1)$ can act
on the base ${\bf D}^2$. Under the ${\bf Z}_k$ projection,
the $SO(3)= SU(2)$ symmetry is intact 
and only charges which are multiples of $k$ are allowed in the theory.

$\bullet$ Case 3
\bea
\label{3}
\pi_3: {\bf S}^4 \rightarrow {\bf D}^3,\quad 
(x_1, x_2, x_3, x_4, x_5) \rightarrow (x_1, x_2, x_3)
\eea 
The generic fiber is a circle ${\bf S}^1$ and 
the discrete subgroups of $SU(2)=SO(3)$
can act on the base ${\bf D}^3$. Under the ${\bf Z}_k$ projection,
the $U(1)$ symmetry coming from ${\bf S^1}$ is intact. 
We know that the (0, 1) theory has a 
global symmetry $SU(2)_R$ and this cannot be realized because
the $SU(2)$ symmetry is broken by the action of the ${\bf Z}_k$ group.
So this case does not fit our goal.

$\bullet$ Case 4
\bea
\label{4}
\pi_4: {\bf S}^4 \rightarrow {\bf D}^4,\quad 
(x_1, x_2, x_3, x_4, x_5) \rightarrow (x_1, x_2, x_3, x_4)
\eea 
The generic fiber is two points (two roots for $x_5$) and 
the base can be embedded into 
a $(u, v)$-complex space ${\bf C}^2$ 
by letting $u = x_1 + i x_2, v = x_3 + ix_4$. Now we can break 
the isometry $SO(5)$ into $SO(4) \times {\bf Z}_2$. 
A  discrete group $\Gamma$ of $ADE$-type can act on the base
 ${\bf D}^4$. By  embedding of $\Gamma$ into the first factor $SU(2)_\Gamma$
 of $SO(4) = SU(2)_\Gamma \times
SU(2)_R$, $SO(5)$ has further broken into $\Gamma \times SU(2)_R$. 
This
${\bf Z}_2$ symmetry interchanges the upper hemisphere and the 
lower hemisphere.

In all four cases we have to be careful with the singular points.
The orbifold acts geometrically on the bases and the origin of the disc is
always a fixed point of the orbifold group. Supergravity alone cannot
describe this fixed point. We would like to emphasize that we need to
include all the degrees of freedom of M theory and not just the eleven
dimensional supergravity ones in order to describe this fixed point.
In addition to untwisted sectors,
M theory includes twisted sectors and these will help to regulate
the singular geometry at the origin of the disk.  Another singularity
appears on the boundary of the disc where the radius of the sphere fiber goes 
to zero. In this case, the fact that we use all the degrees of freedom of M 
theory again helps to regulate the singular geometry.
Besides, the ${\bf S^{4}}$ radius 
$\rho$ is
proportional with $N$ which implies that the disc radii are also proportional 
with $N$. So in the large $N$ limit we can take the boundary arbitrarily 
far away. Therefore the origin and boundary singularities do not appear 
in the full M theory. In our paper we are obtaining the spectrum of chiral
primary operators obtained from the untwisted sector of M theory.
It should be very interesting to obtaine the 
information coming from the twisted sectors of M theory.
     
We are going to work out the cases 2 and 4.
First let us consider the case 4.
We will exhibit the group actions $\Gamma$ in details for the $\pi_4$ case:
\bea
&A_{k-1}(\Gamma ={\bf Z}_k ):& 
u \rightarrow \exp \left(\frac{2\pi i}{k}\right) u, \quad v \rightarrow 
\exp \left(\frac{-2\pi i}{k}\right) v. \\
&D_{k-1}(\Gamma = {\cal D}_{k-2}):&
u \rightarrow \exp \left(\frac{\pi i}{k-2}\right) u, 
\quad v \rightarrow \exp \left(\frac{-\pi i}{k-2}\right) v,\nonumber\\
& & u \rightarrow v,\quad v\rightarrow -u.\nonumber\\
&E_6(\Gamma = {\cal T}):& {\mbox{binary tetrahedral group of order $24$.}}
\nonumber\\
&E_7(\Gamma = {\cal O}):&
 {\mbox{binary octahedral group of order $48$.}}\nonumber \\
&E_8(\Gamma = {\cal I}):&
 {\mbox{binary icosahedral group of order $120$.}}\nonumber
\eea
The Kaluza-Klein harmonics of M theory on
$AdS_{7} \times {\bf S^4}/ \Gamma$ are $\Gamma$ invariant states
and can be obtained by $\Gamma$ projection of the Kaluza-Klein harmonics
on $ AdS_{7} \times {\bf S^4} $ discussed in previous section.
Let us consider the scalar modes with mass formula given by  
(\ref{scalarmass}) and check the relevant and marginal chiral primary
operators in this family. We list  $Sp(4) \rightarrow SU(2)_\Gamma
 \times SU(2)_R$ 
and $Sp(4) \rightarrow SU(2)_R $ branching rules for convenience in the 
Appendix A and B.

$\bullet  A_{k-1}(\Gamma ={\bf Z}_k )$

Before discussing the surviving Kaluza-Klein modes, we briefly discuss 
the SCFT theory obtained on the M5 branes world-volume.
As discuss in \cite{fkpz}, we consider $N$ M5 branes in a transverse space
${\bf R \times C^2/Z_k }$. In the type IIA side this representation
is given by $k$ D6 branes and $N$ NS5 branes. 
The 6 dimensional world-volume theory
is the ${\cal N} = 1, SU(k)^{N-1}$ gauge theory with one tensor
multiplet and $k$ hypermultiplets in the ${\bf k \oplus \bar{k}}$
representation for each of the gauge group factors. Let us 
denote by $U_{a}^{i}, (i = 1, \cdots, N-1; a = 1, \cdots, k)$ the 
hypermultiplets. The field theory which is obtained is conformal.  
The ${\cal N} = 1$ theory in 6 dimensions has a $SU(2)_{R}$
R-symmetry. The scalars in the hypermultiplet transform as a ${\bf 2}$
under $SU(2)_{R}$. The scalars in the vector multiplet
transform as a $\bf 3$ under $SU(2)_{R}$. 
We assign dimension 1 to all of them. 

Now we proceed to discuss the surviving Kaluza-Klein modes.

We first note that  $SU(2)_\Gamma$ representations can be 
realized as representations
on the space of homogeneous polynomials in $u, v$
 and the invariant polynomials 
under the action of ${\bf Z}_k$ are generated by  $uv, u^k, v^k$.
Thus, for general $k$, 
 we will keep only the odd dimensional representations of 
$SU(2)_\Gamma$ since these contain the powers of $uv$.

$i)$
The $k=1$ Kaluza-Klein particle in (\ref{scalarmass}) transforms in the 
${ \bf 5}$ of $SO(5)$. By decomposing ${\bf 5}$ into representations of
$SU(2)_\Gamma \times SU(2)_R$ \cite{ps,mp}, we obtain
\bea
{\bf 5 = (2, 2) \oplus (1, 1)}.
\label{5}
\eea
Now we perform the $\Gamma$ projection on (\ref{5}).
The ${\bf (1, 1) }$ is invariant under the $\Gamma$ so the state
${\bf 1}$ with respect to $SU(2)_R$ will survive the projection.
However, these Kaluza-Klein modes can couple to dimension 2 operators.
Only some of allowed projected states will be survived in the lower
supersymmetry multiplets.
A $(0, 1)$ superconformal subalgebra of $(0, 2)$ superconformal
algebra has a generator, $R$, of the $SU(2)_R$ symmetry.  
The dimensions of chiral operators are determined by $SU(2)_R$ symmetry
half integer, so called spin $s$ and by $SO(6)$ Lorentz quantum numbers.
In this case $SO(6)$ weights are zero. Then the scaling dimensions
of chiral primary operators are determined in terms of the spin as  
\cite{minwalla1}.
\bea
\Delta = 4s
\label{spin}
\eea
This implies that
a dimension 2 chiral primary operator has the spin $1/2$. Therefore
${\bf 1 }$ has the wrong spin to couple a dimension 2 chiral
operator and we do not expect any dimension 2 chiral operators in the 
boundary $(0, 1)$ SCFT. Rather we would expect that 
the dimension 2 operators which
are not chiral primary to couple to the Kaluza-Klein modes in the 
${\bf 1 }$. In the $(0, 2)$ SCFT, ${\bf (2, 2) }$ and ${\bf (1, 1)}$
in (\ref{5}) sit in the same supermultiplet and the masses of 
${ \bf (1, 1 ) }$ Kaluza-Klein states were protected while for $(0, 1)$
SCFT, there is no such guarantee. 

$ii)$
The $k=2$ Kaluza-Klein particle in (\ref{scalarmass}) which transforms
in the ${\bf 14}$ of $SO(5)$ should couple to a dimension 4 chiral primary
operator. Decomposing the ${\bf 14}$ into representations of 
$SU(2)_\Gamma \times SU(2)_R$ gives
\bea
{\bf 14 = (3, 3) \oplus (2, 2) \oplus (1, 1)}.
\label{14}
\eea
The ${\bf (3, 3)}$ and ${\bf (1, 1)}$ are invariants under the
${\bf Z}_k$ projection which give ${\bf 3}$ and ${\bf 1}$ under $SU(2)_R$. 
However, ${\bf 3}$ is the only part which come from the
decomposition (\ref{14}) with the correct spin  to couple a dimension 4
chiral primary operator according to (\ref{spin}). 
Therefore, we expect to have 3 dimension 4 chiral primary operators 
in the boundary SCFT. They are obtained by taking the gauge invariant 
operator $\mbox{Tr}(U^{i_1} U^{i_{2}} U^{i_{3}} U^{j_{4}})$
which is symmetric in the $i$'s indices and antisymmetric in $j$ and either
one of $i$'s. This is just in order to extract the {\bf 3} representation
out of ${\bf 2} \times {\bf 2} \times {\bf 2} \times {\bf 2}$. The $U^{i}$ are
hypermultiplets transforming under consecutive gauge groups in order to obtain
invariants.

$iii)$
The $k=3$ massless Kaluza-Klein particle in (\ref{scalarmass})
should couple to dimension 6 chiral primary operators. It transforms
in the ${\bf 30}$ which decompose as
\bea
{\bf 30 = (4, 4) \oplus (3, 3) \oplus (2, 2) \oplus (1, 1)}.
\eea
The ${\bf (3, 3)}$ and ${\bf (1, 1)}$ are invariant under the
$\Gamma$ projection. However, their spins are not consistent
with coupling to a dimension 6 chiral operator from the analysis of
(\ref{spin}).
Therefore, there are no Kaluza-Klein harmonics in this family
that can couple to dimension 6 chiral primary operators and no such
operators are expected in the boundary $(0, 1)$ SCFT.

$iv)$
The massless vector $k=1$ in (\ref{vectormass}) is in the ${\bf 10}$
of $SO(5)$. Decomposing ${\bf 10}$ we get
\bea
{\bf 10  = (3, 1) \oplus (2, 2) \oplus (1, 3)}.
\label{10}
\eea
The ${\bf (3, 1)}$ and ${\bf (1, 3)}$ are invariant under the
$\Gamma$ projection. From the result of \cite{minwalla1}, isolated 
values of dimensions occur at $\Delta=4s + 5$ for the vector. This implies that
${\bf (3, 1)}$ is the only part in the
decomposition (\ref{10}) with the correct spin  to couple to a dimension 5
chiral primary operator.
Therefore, we expect the superconformal primary to be 
given by a  ${\bf 1}$ i.e. a singlet of $SU(2)_R$. In terms of the operator 
spectrum of the six dimensional SCFT a gauge invariant combination
having dimension 5 and being a singlet of $SU(2)_R$
is $\mbox{Tr}(U \Phi U U^2)$ which is antisymmetric in the last two $U$'s. 

It is rather straightforward to check the scaling dimensions with right 
quantum numbers for other massive representations: spinors, higher rank
tensors and so on. 

Now we turn to the binary dihedral group.

$\bullet  D_{k-1}(\Gamma ={\cal D}_{k-2} ), E_n$

Since the binary dihedral group ${\cal D}_{k-2}$ is generated by
the cyclic group 
${\bf  Z}_{k-2/2}$ and a reflection, it will further restrict the 
representations. For our purpose, notice that ${\bf 3}$ does not have an 
invariant
component since $uv$ is not invariant under the reflection.
However, ${\bf 1}$ has an invariant component for $ D_{k-1}$.
Again the 6 dimensional field theory brane configuration \cite{fkpz} 
is the same as the one
for $A_{k-1}$ gauge theory except that there exists an orientifold 6 plane.
Each gauge group factors couple hypermultiplets in vetor representation or
fundamental representation.
We expect the superconformal primary to be 
given by a  ${\bf 1}$ i.e. a singlet of $SU(2)_R$. In terms of the operator 
spectrum of the SCFT a gauge invariant combination
having dimension 5 and being a singlet of $SU(2)_R$
is $\mbox{Tr}(U \Phi U U^2)$ which is antisymmetric in the last two $U$'s. 
Since ${\cal D}_2$ is the subgroup of $E_6$ and $E_7$, ${\bf 3}$ does not
have an invariant component under these groups. Similary, one can see
that $E_8$ does not leave any invariant subspace of ${\bf 3}$.
It is true that ${\bf 1}$ has an invariant component for $E_n$.
But it is not clear how this can be realized in the SCFT side because there 
is no known field theory for $E_n$ gauge theory.

Now we go to the case 2 i.e. to the projection $\pi_2$ as in
(\ref{2}). In this case the $SO(5)$ group is broken to
$SU(2)_R \times U(1)/{\bf Z}_k$. The branching rule and (\ref{spin}) analysis  
tells us that there is no allowed $\Gamma$ invariant states.


\section{ Conclusion }
\setcounter{equation}{0}

We studied the relation between chiral primary operators of
SCFT in 6 dimensions and the KK states of M theory on orbifolds of
$AdS_7 \times {\bf S^4}$. This generalizes the relation between the chiral
primary operators of $(0, 2)$ SCFT and KK states of supergravity on
$AdS_7 \times {\bf S^4}$ found in \cite{aoy,lr,minwalla2,halyo1}. In our case,
we have obtained 
the surviving KK modes in the orbifold models by projecting on
$\Gamma$ invariant states.
the known KK states for supergravity
compactified on $AdS_7 \times {\bf S^4}$ .

\vspace{2cm}
\centerline{\bf Acknowledgments} 

We thank J. Gomis, M. Gunaydin, A. Kehagias, S. Minwalla, 
Y. Oz, J. Terning and A. Zaffaroni for email 
correspondences.
RT would like to thank Prof. Orlando Alvarez for  valuable
discussions on various points in this paper.
CA thanks B.-H. Lee for bring his attention to the ref. \cite{ps}. 

We would like to thank our referee for helping us to improve our paper.

\section{Appendix A: $Sp(4)$ Branching Rule I}

\begin{tabular}[b]{|c|l|l|}
\hline Fields  & $Sp(4)$ Dynkin label & $SU(2)_\Gamma \times SU(2)_R$ \\ \hline
$\sqrt{  \mbox {3 index anti. tensor}}$  
& $(0, 0): \;\;\; {\bf 1}$    & ${ \bf (1, 1) }$  \\ \hline
 spinor  & $(1, 0): \;\;\; {\bf 4}$  
              & ${ \bf (2, 1) \oplus (1, 2) }$  \\ \hline
 ${\mbox {scalar}}^{*}$ 
& $(0, 1): \;\;\; {\bf 5}$ & ${ \bf (2, 2) \oplus (1, 1)}$ \\ \hline
\hline 
graviton & $(0, 0): \;\;\; {\bf 1}$    & ${ \bf (1, 1) }$  \\ \hline
 gravitino & $(1, 0): \;\;\; {\bf 4}$  
              & ${ \bf (2, 1) \oplus (1, 2) }$  \\ \hline
 $\sqrt{  \mbox { 3 index anti. tensor}}$ 
& $(0, 1): \;\;\; {\bf 5}$ & ${ \bf (2, 2) \oplus (1, 1)}$ \\ \hline
${\mbox {vector}}^{*}$   
& $(2, 0): \;\;\; {\bf 10}$ 
& $ { \bf {(3, 1)}^{\#} \oplus (2, 2) \oplus (1, 3)}$ \\ \hline
 $ {\mbox {scalar}}^{*}$ & 
$(0, 2): \;\;\; {\bf 14}$ & $ { \bf {(3, 3)}^{\#} \oplus (2, 2) \oplus 
(1, 1)}$ \\ \hline
 spinor & $(1, 1): \;\;\; {\bf 16}$ & $ { \bf (3, 2) \oplus (2, 3) 
\oplus (2, 1) \oplus (1, 2)}$ \\ 
\hline
\hline
$\sqrt{  \mbox {3 index anti. tensor}}$ 
& $(0, 0): \;\;\; {\bf 1}$    & ${ \bf (1, 1) }$  \\ \hline
 gravitino & $(1, 0): \;\;\; {\bf 4}$  
              & ${ \bf (2, 1) \oplus (1, 2) }$  \\ \hline
 graviton & $(0, 1): \;\;\; {\bf 5}$ & ${ \bf (2, 2) \oplus (1, 1)}$ \\ \hline
 2 index anti. tensor   
& $(2, 0): \;\;\; {\bf 10}$ 
& $ { \bf (3, 1) \oplus (2, 2) \oplus (1, 3)}$ \\ \hline
 $\sqrt{  \mbox {3 index anti. tensor}}$ 
& $(0, 2): \;\;\; {\bf 14}$ & $ { \bf (3, 3) \oplus (2, 2) \oplus 
(1, 1)}$ \\ \hline
 gravitino & $(1, 1): \;\;\; {\bf 16}$ & $ { \bf (3, 2) \oplus (2, 3) 
\oplus (2, 1) \oplus (1, 2)}$ \\ 
\hline
 spinor & $(3, 0): \;\;\; {\bf 20}$ & $ { \bf (4, 1) \oplus (3, 2) 
\oplus (2, 3) \oplus 
(1, 4)}$ \\ 
\hline
$ {\mbox {scalar}}^{*}$ & $(0, 3): \;\;\; {\bf 30}$ 
& $ { \bf (4, 4) \oplus (3, 3) 
\oplus (2, 2) \oplus (1, 1)} $
\\ \hline
vector & $(2, 1): \;\;\; {\bf 35}$ &$ {\bf (4, 2) 
\oplus (3, 3) \oplus 
(2, 4) \oplus 
(3, 1)  }$\\
 & & ${ \bf \oplus (2, 2) \oplus (1, 3)}$ \\ \hline
spinor & $(1, 2): \;\;\; {\bf 40}$ 
& ${ \bf (4, 3) \oplus (3, 4) \oplus (3, 2) \oplus (2, 3)
}$ \\ 
& & $ { \bf \oplus (2, 1) \oplus (1, 2) } $\\ \hline
\end{tabular}

Table 1. The first 3 superconformal multiplets, their
$Sp(4)$ Dynkin labelling and branching rules for $SU(2)_\Gamma
 \times SU(2)_R$. 
The asterisk $*$ denote the spectrums of negative or zero $m^2$.
The two states with $\#$ are $\Gamma$ invariant ones with appropriate
right quantum number.
 
\begin{tabular}[b]{|c|l|l|}
\hline Fields  & $Sp(4)$ Dynkin label & $SU(2)_\Gamma \times SU(2)_R$ \\ \hline
scalar  & $(0, 0): \;\;\; {\bf 1}$    & ${ \bf (1, 1) }$  \\ \hline
 spinor  & $(1, 0): \;\;\; {\bf 4}$  
              & ${ \bf (2, 1) \oplus (1, 2) }$  \\ \hline
 $\sqrt{  \mbox { 3 index anti. tensor}}$  
 & $(0, 1): \;\;\; {\bf 5}$ & ${ \bf (2, 2) \oplus (1, 1)}$ \\ \hline
vector 
& $(2, 0): \;\;\; {\bf 10}$ 
& $ { \bf (3, 1) \oplus (2, 2) \oplus (1, 3)}$ \\ \hline
 graviton & $(0, 2): \;\;\; {\bf 14}$ & $ { \bf (3, 3) \oplus (2, 2) \oplus 
(1, 1)}$ \\ \hline
 gravitino & $(1, 1): \;\;\; {\bf 16}$ & $ { \bf (3, 2) \oplus (2, 3) 
\oplus (2, 1) \oplus (1, 2)}$ \\ 
\hline
 spinor & $(3, 0): \;\;\; {\bf 20}$ & $ { \bf (4, 1) \oplus (3, 2) 
\oplus (2, 3) \oplus 
(1, 4)}$ \\ 
\hline
 $\sqrt{  \mbox {3 index anti. tensor}}$ & $(0, 3): \;\;\; {\bf 30}$ 
& $ { \bf (4, 4) \oplus (3, 3) 
\oplus (2, 2) \oplus (1, 1)} $
\\ \hline
2 index anti. tensor & $(2, 1): \;\;\; {\bf 35}$ &$ {\bf (4, 2) 
\oplus (3, 3) \oplus 
(2, 4) \oplus 
(3, 1)  }$\\
 & & ${ \bf \oplus (2, 2) \oplus (1, 3)}$ \\ \hline
scalar & $(4, 0): \;\;\; {\bf 35}$ 
& ${ \bf (5, 1) \oplus (4, 2) \oplus (3, 3) \oplus (2, 4) 
 }$ \\  
& & ${\bf \oplus (1, 5)} $\\ \hline
gravitino & $(1, 2): \;\;\; {\bf 40}$ 
& ${ \bf (4, 3) \oplus (3, 4) \oplus (3, 2) \oplus (2, 3)
}$ \\ 
& & $ { \bf \oplus (2, 1) \oplus (1, 2) } $\\ \hline
 scalar & $(0, 4): \;\;\; {\bf 55}$  & ${ \bf (5, 5) \oplus (4, 4) 
\oplus (3, 3) \oplus (2, 2)
}$ \\ 
& & ${\bf \oplus (1, 1) }$ \\ \hline
spinor & $(3, 1): \;\;\; {\bf 64}$ 
& $ { \bf (5, 2) \oplus (4, 3) \oplus (3, 4) \oplus (2, 5) 
} $ \\ 
& & ${ \bf \oplus
 (4, 1) \oplus (3, 2) \oplus (2, 3) \oplus (1, 4)}$ \\ \hline
spinor & $(1, 3): \;\;\; {\bf 80 } $ 
& ${ \bf (5, 4) \oplus (4, 5) \oplus (4, 3) \oplus (3, 4) 
}$ \\ 
& & ${\bf \oplus (3, 2) \oplus (2, 3) \oplus (2, 1) \oplus (1, 2) }$ 
\\ \hline
vector & $(2, 2): \;\;\; {\bf 81}$ 
& ${ \bf (5, 3) \oplus (4, 4) \oplus (3, 5) \oplus (4, 2)  
}$ \\ 
& & ${\bf \oplus (3, 3)  \oplus (2, 4) \oplus (3, 1) 
\oplus (2, 2) \oplus (1, 3)}$ \\ \hline
\end{tabular}

Table 2. The fourth superconformal multiplet, its
$Sp(4)$ Dynkin labelling and branching rules for 
$SU(2)_\Gamma \times SU(2)_R$.  

\section{Appendix B: $Sp(4) $ Branching Rule II}

\begin{tabular}[b]{|c|l|l|}
\hline Fields  & $Sp(4)$ Dynkin label & $SU(2)_R$ \\ \hline
$\sqrt{  \mbox {3 index anti. tensor}}$
& $(0, 0): \;\;\; {\bf 1}$    & ${ \bf 1}$  \\ \hline
 spinor  & $(1, 0): \;\;\; {\bf 4}$
              & ${ \bf 4 }$  \\ \hline
 ${\mbox {scalar}}^{*}$
& $(0, 1): \;\;\; {\bf 5}$ & ${ \bf 5 }$ \\
\hline
\hline
graviton & $(0, 0): \;\;\; {\bf 1}$    & ${ \bf 1 }$  \\ \hline
 gravitino & $(1, 0): \;\;\; {\bf 4}$
              & ${ \bf 4 }$  \\ \hline
 $\sqrt{  \mbox { 3 index anti. tensor}}$
& $(0, 1): \;\;\; {\bf 5}$ & ${ \bf 5}$ \\ 
\hline
\hline
${\mbox {vector}}^{*}$
& $(2, 0): \;\;\; {\bf 10}$
& $ { \bf 7 \oplus 3}$ \\ \hline
 $ {\mbox {scalar}}^{*}$ &
$(0, 2): \;\;\; {\bf 14}$ & $ { \bf 9 \oplus 5}$ \\ \hline
spinor & $(1, 1): \;\;\; {\bf 16}$ & $ { \bf 8
\oplus 6 \oplus 2}$ \\
\hline
\hline
$\sqrt{  \mbox {3 index anti. tensor}}$
& $(0, 0): \;\;\; {\bf 1}$    & ${ \bf 1}$  \\ \hline
 gravitino & $(1, 0): \;\;\; {\bf 4}$
              & ${ \bf 4}$  \\ \hline
 graviton & $(0, 1): \;\;\; {\bf 5}$ & ${ \bf 5}$ \\ \hline
 2 index anti. tensor
& $(2, 0): \;\;\; {\bf 10}$
& $ { \bf 7 \oplus 3}$ \\ \hline
 $\sqrt{  \mbox {3 index anti. tensor}}$
& $(0, 2): \;\;\; {\bf 14}$ & $ { \bf 9 \oplus 5}$ \\ \hline
 gravitino & $(1, 1): \;\;\; {\bf 16}$ & $ { \bf 8 \oplus 6
\oplus 2}$ \\
\hline
 spinor & $(3, 0): \;\;\; {\bf 20}$ & $ { \bf 10 \oplus 6
\oplus 4}$ \\
\hline
 ${\mbox {scalar}}^{*}$ & $(0, 3): \;\;\; {\bf 30}$
& $ { \bf 13 \oplus 9
\oplus 7  \oplus 1} $ \\ 
 \hline
vector & $(2, 1): \;\;\; {\bf 35}$ &$ {\bf 11
\oplus 9  \oplus 7 \oplus 5 \oplus 3 }$\\
 \hline
spinor & $(1, 2): \;\;\; {\bf 40}$
& ${ \bf 12 \oplus 10  \oplus 8 \oplus 6 \oplus 4}$ \\
\hline
\end{tabular}

Table 3. The first 3 superconformal multiplets, their
$Sp(4)$ Dynkin labelling and branching rules for $SU(2)_R$.
The asterisk $*$ denote the spectrums of negative or zero $m^2$.

\begin{tabular}[b]{|c|l|l|}
\hline Fields  & $Sp(4)$ Dynkin label & $SU(2)_R$ \\ \hline
scalar  & $(0, 0): \;\;\; {\bf 1}$    & ${ \bf 1 }$  \\ \hline
 spinor  & $(1, 0): \;\;\; {\bf 4}$  
              & ${ \bf 4}$  \\ \hline
 $\sqrt{  \mbox { 3 index anti. tensor}}$  
 & $(0, 1): \;\;\; {\bf 5}$ & ${ \bf 5 }$ \\ \hline
vector 
& $(2, 0): \;\;\; {\bf 10}$ 
& $ { \bf 7 \oplus 3}$ \\ \hline
 graviton & $(0, 2): \;\;\; {\bf 14}$ & $ { \bf 9 \oplus 5 }$ \\ \hline
 gravitino & $(1, 1): \;\;\; {\bf 16}$ & $ { \bf 8 \oplus 6 
\oplus 2 }$ \\ 
\hline
 spinor & $(3, 0): \;\;\; {\bf 20}$ & $ { \bf 10 \oplus 6 
\oplus 4 }$ \\ 
\hline
 $\sqrt{  \mbox {3 index anti. tensor}}$ & $(0, 3): \;\;\; {\bf 30}$ 
& $ { \bf 13 \oplus 9 
\oplus 7 \oplus 1} $
\\ \hline
2 index anti. tensor & $(2, 1): \;\;\; {\bf 35}$ &$ {\bf 11 
\oplus9 \oplus 7 \oplus 
5 \oplus 3 }$ \\ \hline
scalar & $(4, 0): \;\;\; {\bf 35}$ 
& ${ \bf 13 \oplus 9 \oplus 7 \oplus 5 \oplus 1 
 }$ \\  \hline
gravitino & $(1, 2): \;\;\; {\bf 40}$ 
& ${ \bf 12 \oplus 10 \oplus 8 \oplus 6 \oplus 4 } $ \\  \hline
 scalar & $(0, 4): \;\;\; {\bf 55}$  & ${ \bf 17 \oplus 13 
\oplus 11 \oplus 9 \oplus 5 }$ \\ \hline
spinor & $(3, 1): \;\;\; {\bf 64}$ 
& $ { \bf 14 \oplus 12 \oplus 10 \oplus 8 
} $ \\ 
& & ${ \bf \oplus
 8 \oplus 6 \oplus 4 \oplus 2}$ \\ \hline
spinor & $(1, 3): \;\;\; {\bf 80 } $ 
& ${ \bf 16 \oplus 14 \oplus 12 \oplus 10 
}$ \\ 
& & ${\bf \oplus 10 \oplus 8 \oplus 6 \oplus 4 }$ 
\\ \hline
vector & $(2, 2): \;\;\; {\bf 81}$ 
& ${ \bf 15 \oplus 13 \oplus 11 \oplus 11  }$ \\ 
& & ${\bf \oplus 9  \oplus 7 \oplus 7 
\oplus 5 \oplus 3}$ \\ \hline
\end{tabular}

Table 4. The fourth superconformal multiplet, its
$Sp(4)$ Dynkin labelling and branching rules for $SU(2)_R$.


\end{document}